\documentclass[journal=jacsat,
manuscript=article,
]{achemso}

\usepackage{chemformula} 
\usepackage[T1]{fontenc} 
\usepackage{mathrsfs,amsmath,amssymb,bm}
\usepackage{graphicx}
\usepackage{subfigure}
\usepackage[ruled,linesnumbered]{algorithm2e}
\usepackage{multirow}
\usepackage{booktabs}
\SectionsOn

\newcommand{\br}{\bm{r}}
\newcommand{\bu}{\mathbf{u}}

\newcommand{\bR}{\bm{R}}

\newcommand{\bS}{\boldsymbol{S}}

\newcommand{\bM}{\mathbf{M}}
\newcommand{\bI}{\boldsymbol{I}}
\newcommand{\Rmnum}[1]{\expandafter\@slowromancap\romannumeral #1@}

\usepackage { xr }

\SectionNumbersOn

\author{Zhijuan He}
\affiliation[Xiangtan University]
{Hunan Key Laboratory for Computation and Simulation in Science and Engineering, Key Laboratory of Intelligent Computing and Information Processing of Ministry of Education, School of Mathematics and Computational Science, Xiangtan University, Xiangtan, Hunan, 411105,China}
\author{Xin Wang}
\affiliation[Xiangtan University]
{Hunan Key Laboratory for Computation and Simulation in Science and Engineering, Key Laboratory of Intelligent Computing and Information Processing of Ministry of Education, School of Mathematics and Computational Science, Xiangtan University, Xiangtan, Hunan, 411105,China}

\author{Pingwen Zhang}
\email{pzhang@pku.edu.cn}
\affiliation[Wuhan UniVersity and Peking University]
{School of Mathematics and Statistics, Wuhan University, Wuhan,  430072, China. School of Mathematical Sciences, Peking University, Beijing, 100871, China}

\author{An-Chang Shi}
\email{shi@mcmaster.ca}
\affiliation[McMaster UniVersity]
{Department of Physics and Astronomy, McMaster University, 
	Hamilton, Ontario L8S 4M1, Canada}
\author{Kai Jiang}
\affiliation[Xiangtan University]
{Hunan Key Laboratory for Computation and Simulation in Science and Engineering, Key Laboratory of Intelligent Computing and Information Processing of Ministry of Education, School of Mathematics and Computational Science, Xiangtan University, Xiangtan, Hunan, 411105,China}
\email{kaijiang@xtu.edu.cn}

\title[An \textsf{achemso} demo]
  {Theory of polygonal phases self-assembled from T-shaped liquid crystalline molecules}

\abbreviations{IR,NMR,UV}

\begin{document}



\begin{abstract}
Extensive experimental studies have shown that numerous ordered phases can be formed via the self-assembly of T-shaped liquid crystalline molecules (TLCMs) composed of a rigid backbone, two flexible end chains and a flexible side chain. However, a comprehensive understanding of the stability and formation mechanisms of these intricately nanostructured phases remains incomplete. Here we fill this gap by carrying out a theoretical study of the phase behaviour of TLCMs. Specifically, we construct phase diagrams of TLCMs by computing the free energy of different ordered phases of the system. 
Our results reveal that the number of polygonal edges increases as the length of side chain or interaction strength increases, consistent with experimental observations. 
The theoretical study not only reproduces the experimentally observed phases and phase transition sequences, but also systematically analyzes the stability mechanism of the polygonal phases. 

\end{abstract}


\section{Introduction}
\label{sec:intrd}
Liquid crystalline  molecules (LCMs) are a class of soft materials that can self-assemble into numerous ordered structures 
in both crystalline states and liquid crystalline states\,\cite{lyu2020liquid}. The rich phase behaviour and unique properties of LCMs make them useful advanced materials with applications in many fields such as biomedical engineering, electronics and communications\,\cite{woltman2007liquid,larsen2003optical, wang2016stimuli, yang2021beyond}. Among the many types of LCMs, the T-shaped liquid crystalline molecules (TLCMs), composed of a rigid backbone with two incompatible end blocks and a flexible side chain, have been extensively studied experimentally\,\cite{koelbel2001design, cheng2003calamitic, cheng2004liquld,  chen2005liquid, chen2005carbohydrate, tschierske2007liquid, liu2007triangle,  liu2008trapezoidal, cheng2011influence, zeng2011complex, tschierske2013development,  lehmann2018soft, alexander2020liquid}, revealing that these LCMs can self-assemble into an amazing array of complex ordered phases. Specifically, increasing the side chain length results in an interesting phase transition sequence of one-dimensional smectic phases $\rightarrow$ simple polygons $\rightarrow$ giant polygons $\rightarrow$ three-dimensional lamellar phases $\rightarrow$ three-dimensional bicontinuous cubic networks. Besides providing a platform to engineering intricately nanostructured materials, the TLCMs offer an interesting model system to study the self-assembly and stability of complex ordered phases from macromolecular systems containing rigid and flexible components.

Theoretical and simulation studies can provide a good understanding of the phase behaviour of self-assembling macromolecules.
For the case of TLCMs, several simulation methods have been used to investigate their self-assembly. 
Specially, molecular dynamics of coarse-grained models has been employed to explore the phase behaviours of several TLCM systems\,\cite{ crane2010lyotropic, crane2011selfassembly, sun2017molecular, fayaztorshizi2021coarse}.
These studies observed the formation of layered phases, simple polygons, and three-dimensional bicontinuous cubic networks. 
Furthermore, dissipative particle dynamics simulations\,\cite{ bates2009dissipative, bates2010computer, bates2011computer, liu2013dissipative}
have been utilized to study the influence of side chain length, temperature, and hydrogen bonding on the phase behaviour of TLCMs,
and observed layered, simple polygonal, giant polygonal and gyroid phases.
Monte Carlo simulations\,\cite{peroukidis2012entropy} have been used to study the role of entropy player in self-assembled layered and hexagonal phases. 
These simulation studies of TLCMs mainly looked at the layered phases and simple polygons, with limited results of giant polygons. 
A comprehensive understanding of the stability and formation mechanisms of these intricately liquid crystalline polygons remains  incomplete. 

In this work, we report a theoretical study of the phase behaviour of TLCMs by using the self-consistent field theory (SCFT), which is a flexible and powerful theoretical framework for analyzing the equilibrium phase behaviours of inhomogeneous macromolecular systems. It has been successfully applied to flexible and semi-flexible polymeric systems\,\cite{morse1994semiflexible, masten1998liquid, Song2009New, gao2011self-assembly,  jiang2013influence, zhu2013selfassembly, song2011phase, cai2017liquild, liu2018archimedean}.
In our SCFT study of the TLCMs, the rigid liquid crystalline segments are described as worm-like chains with liquid crystalline interactions, whereas the end- and side-chains are modelled as flexible Gaussian chains. The resulting SCFT equations represent a great computational challenge due to the existence of both flexible and semi-flexible components. 
We overcome this challenge by developing an efficient and precise parallel algorithm to solve the SCFT equations, 
enabling us to obtain solutions corresponding to many polygonal and layered phases.
The thermodynamic stability of these ordered structures is examined by comparing their free energy. Phase diagrams of the system are constructed in the plane spanned by the volume fraction of side chain and interaction strength. 
Furthermore, we model the experimental process by changing the number of side chain monomers and obtain a phase transition sequence that is consistent with experimental observations.

\section{Model and Methods}
\begin{figure}[!htbp]
	\centering
	\includegraphics[width=6cm]{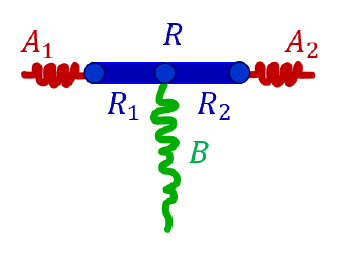}
	\caption{Schematic of TLCM chain containing a rigid backbone block $R$ (blue),
		two ends flexible blocks $A$ (red),
		and a flexible side block $B$ (green).}
	\label{fig:model}
\end{figure}

We consider an incompressible melt consisting of $n$ TLCMs in a volume $V$.
Each TLCM, with a degree of polymerization $N$, consists of five blocks constructed from three chemically distinct monomers ($A,\, B,\, R$), as shown schematically in Fig.\,\ref{fig:model}.
The number of monomers for the five blocks is denoted by $N_{i}=f_{i}N$, where $f_{i}$ is the volume fraction of the $i$ block, $i=A_{1},\, A_{2},\, B,\, R_{1},\,R_{2}$. It is noted that $f_{A_{1}}+f_{A_{2}}+f_{B}+f_{R_{1}}+f_{R_{2}}=1$, $N_{A_{1}}+N_{A_{2}}+N_{B}+N_{R_{1}}+N_{R_{2}}=N$. 
The statistical segment lengths of monomers $\alpha$ are $b_{\alpha}$ ($\alpha \in \{A,\,B,\, R \}$), respectively. 
We employ the Gaussian chain model and the wormlike chain model to describe
flexible and semi-flexible blocks, respectively\,\cite{fredrickson2006equilibrium}.
The conformation of a block is described by a space curve $\bR^{i}_{\alpha}(s)\,(s\in I_{\alpha})$, where $I_{A}=I_{A_{1}}\cup I_{A_{2}}=[0,f_{A_{1}}]\cup [0,f_{A_{2}}]$, $I_{B}=[0,f_{B}]$, and $I_{R}= I_{R_{1}}\cup I_{R_{2}}= [f_{A_{1}}, f_{A_{1}}+f_{R_{1}}]\cup [f_{A_{2}}, f_{A_{2}}+f_{R_{2}}]$, which specifies the position of the $s$-th monomer in the $\alpha$-block of the $i$-th chain. 
According to this definition, the normalized concentrations of monomers $A,\,B$ and $R$ at a spatial position $\br$ are
\begin{equation}
	\begin{split}
		\hat{\rho}_{A}({\br})&=
		\frac{N}{\rho_{0}}\sum_{i=1}^{n}
		\int_{I_{A}} \delta\left[{\br}-\bR^{i}_{A}(s)\right] \mathrm{d} s,
		\\
		\hat{\rho}_{B}({\br})&=\frac{N}{\rho_{0}}\sum_{i=1}^{n} 
		\int_{I_{B}} \delta\left[{\br}-\bR^{i}_{B}(s)\right] \mathrm{d} s ,\\
		\hat{\rho}_{R}({\br})
		&=\frac{N}{\rho_{0}}\sum_{i=1}^{n}
		\int_{I_{R}}\delta\left [\br-\bR^{i}_{R}(s) \right ] \mathrm{d} s,
	\end{split}
\end{equation}
where $\rho_{0}$ is the per unit volume of density.
The incompressibility condition requires $\hat{\rho}_{A}({\br})+\hat{\rho}_{B}({\br})+\hat{\rho}_{R}({\br})=1$.

There are various order parameters to describe the orientational order of rigid segments\,\cite{selinger2016introduction}. Here we choose the two-dimensional order parameter,
\begin{equation}
	\begin{split}
		\hat{\bS}({\br})=&\frac{N}{\rho_{0}}\sum_{i=1}^{n} 
		\int_{I_{R}} \delta\left [\br-\bR^{i}_{R}(s)\right ] \big(\bu^{i}(s)\bu^{i}(s)-\frac{\bI}{2}\big)\mathrm{d} s,
	\end{split}
\end{equation}
where $\bu^{i}(s)=d\bR^{i}_{R}(s)/ds$ is the unit tangent vector to the 
semi-flexible block at contour location $s$.
The stretching conformational energy of non-interacting flexible chains is
\begin{equation}
	\begin{split}
		H_{0}
		=&\frac{3}{2b_A^2}\sum_{i=1}^{n}\int_{I_{A}}\left|\frac{d~\bR^{i}_{A}(s)}{ds}\right|^2 ds 
		+\frac{3}{2b_B^2}\sum_{i=1}^{n}\int_{I_{B}}\bigg|\frac{d~\bR^{i}_{B}(s)}{ds} \bigg|^2 ds.
	\end{split}
\end{equation}
The bending conformational energy of non-interacting semi-flexible blocks is
\begin{equation}
	\begin{split}
		H_{1}
		&=\frac{\lambda}{2b_R^2}\sum_{i=1}^{n}\int_{I_{R}}\left|\frac{d~\bu^{i}(s)}{ds} \right|^2 ds,
	\end{split}
\end{equation}
where $\lambda$ is the stiffness of the semi-flexible block. 
The energy of the parallel alignment between semi-flexible chains, described using the Maier-Saupe type of orientational interaction,
\begin{equation}
	\begin{split}
		H_{S}=-\frac{\eta \rho_{0}}{2}\int\hat{\bS}(\br):\hat{\bS}(\br)~d\br,
	\end{split}
\end{equation}
where Maier-Saupe parameter $\eta$  represents the magnitude of the orientational interaction that favors parallel alignment of the semi-flexible segments.
Following the standard Flory-Huggins approach, the interaction potential $H_{F}$ of the system is given by, 
\begin{equation}
	\begin{split}
		H_{F} = &   \rho_{0} \int\big[ \bar{\chi}_{AA}\hat{\rho}_{A}(\br)\hat{\rho}_{A}(\br) + 2\bar{\chi}_{AB}\hat{\rho}_{A}(\br)\hat{\rho}_{B}(\br)\\
		& + \bar{\chi}_{BB}\hat{\rho}_{B}(\br)\hat{\rho}_{B}(\br)
		+ 2\bar{\chi}_{AR}\hat{\rho}_{A}(\br)\hat{\rho}_{R}(\br)\\
		& + 2\bar{\chi}_{BR}\hat{\rho}_{B}(\br)\hat{\rho}_{R}(\br)
		+\bar{\chi}_{RR}\hat{\rho}_{R}(\br)\hat{\rho}_{R}(\br)
		\big]\,d \br,
	\end{split}
	\label{eq:HF1}
\end{equation}
where the Flory-Huggins interaction parameter $\bar{\chi}_{ij}$ ($i,\,j \in \{ A,\, B,\, R \}$) represents the interaction between monomers $i$ and $j$. It is assumed that these interaction parameters $\bar{\chi}_{ij}$ can be positive or negative, representing repulsive and attractive interactions, respectively. 
Hydrogen bonding is commonly modelled by attractive interaction\,\cite{lee2006hydrogen}, i.e., $\bar{\chi}_{ij}<0$.
Considering the incompressible condition and ignoring the contributions from terms linear in the monomer density, the interaction potential $H_{F}$ becomes
\begin{equation}
	\begin{split}
		H_{F} = & \rho_{0} \int\big[ \chi_{AB}\hat{\rho}_{A}(\br)\hat{\rho}_{B}(\br)\\
		& + \chi_{AR}\hat{\rho}_{A}(\br)\hat{\rho}_{R}(\br)
		+ \chi_{BR}\hat{\rho}_{B}(\br)\hat{\rho}_{R}(\br)
		\big]\,d \br,
	\end{split}
	\label{eq:HF}
\end{equation}
where the effective Flory-Huggins parameters are given by $\chi_{ij}=2\bar{\chi}_{ij}- (\bar{\chi}_{ii}+\bar{\chi}_{jj} )$.

The particle-based partition functional is,
\begin{equation}
	\begin{split}
		\mathcal{Z}=&
		\frac{z^{n}_{T}}{n!}
		\int\int
		\delta [\hat{\rho}_{A}(\br)+\hat{\rho}_{B}(\br)+\hat{\rho}_{R}(\br)-1]\\
		&\exp{[-H_{0}-H_{1}-H_{F}-H_{S}]}~d\bu(s)~\mathcal{D}\bR^{i}_{\alpha}(s),
	\end{split}
	\label{eq:Z}
\end{equation} 
where $z_{T}$ is the partition function of the TLCM chain, which is influenced by kinetic energy. The delta function constrains the local incompressibility condition.
Taking the Hubbard-Stratonovich transformation and the saddle-point approximation\,\cite{fredrickson2006equilibrium}, 
the particle form of the partition function  can be transformed into 
the mean field form of the partition function as
\begin{equation}
	\begin{split}
		\mathcal{Z} \propto \int \int \int \int  \exp\big(-H\big[\mu_{+},\mu_{1},\mu_{2},\bM \big]\big)
		\,\mathcal{D}\mu_{+}\,\mathcal{D}\mu_{1}\,\mathcal{D}\mu_{2}\,\mathcal{D}\bM.
	\end{split}
\end{equation}

The free energy per chain in the unit of thermal energy $k_{B}T$ at temperature $T$, where $k_{B}$ is the Boltzmann constant, can be expressed as
\begin{equation}
	\begin{split}
		\frac{H}{nk_{B}T}=&\frac{1}{V}\int_{\mathcal{B}}	\big(\frac{1}{4N\zeta_{1}}\mu_{1}^2(\br)+\frac{1}{4N\zeta_{2}}\mu_{2}^2(\br)-\mu_{+}(\br)\big)~d\br\\
		&+\frac{1}{2\eta NV}\int_{\mathcal{B}} \bM(\br):\bM(\br)~d\br
		-\log Q,
		\label{H}
	\end{split}
\end{equation}

The single-chain free energy, $H/nk_{B}T$, of the system can be divided
into three parts: interfacial energy $H_{inter}/nk_{B}T$, orientation interaction energy $H_{orien}/nk_{B}T$, and entropic energy $ -TS/nk_{B}T$,

\begin{equation}
	\begin{split}
		\frac{H_{inter}}{nk_{B}T} = & \frac{1}{V}\int_{\mathcal{B}}	 \frac{1}{4N\zeta_{1}}\mu_{1}^2(\br)+\frac{1}{4N\zeta_{2}}\mu_{2}^2(\br)-\mu_{+}(\br)\big)~d\br,\\
		\frac{H_{orien}}{nk_{B}T} = & \frac{1}{2\eta NV}\int_{\mathcal{B}} \bM(\br):\bM(\br)~d\br,\\
		\frac{-TS}{nk_{B}T}  = & -\log Q,
	\end{split}
	\label{eq:energySplit}
\end{equation}
where $\mu_{1}(\br)$,\, $\mu_{2}(\br)$ are general ``exchange'' chemical 
potentials of the system, $\mu_{+}(\br)$ is the ``pressure'' 
chemical potential to ensure the local incompressibility of 
the system, $\bM(\br)$ is the orientation tension field of 
semi-flexible segments, and $Q$ is the single chain partition function. 
The parameters in Eq.\,(\ref{H}) are defined by
\begin{equation}
	\begin{split}
		\zeta_{1}=&\frac{-\Delta}{4\chi_{AB}},\qquad\qquad \zeta_{2}=\chi_{AB},\\
		\Delta=&\chi_{AB}^2+\chi_{AR}^2+\chi_{BR}^2-2\chi_{AB}\chi_{AR}\\
		&-2\chi_{AB}\chi_{BR}-2\chi_{AR}\chi_{BR}.
		\nonumber
	\end{split}
\end{equation}

The mean fields $\omega_{\alpha}(\br)$ ($\alpha\in \{A,\,B,\,R\}$) are the function of $\mu_{+}(\br),\,\mu_{1}(\br),\,\mu_{2}(\br)$,
\begin{equation}
	\omega_{\alpha}(\br)=\mu_{+}(\br)-\sigma_{1\alpha}\mu_{1}(\br)-\sigma_{2\alpha}\mu_{2}(\br),
	\label{eq:w}
\end{equation}
where
\begin{equation}
	\begin{split}
		\sigma_{1A}=&\frac{1}{3},\quad\sigma_{1R}=-\frac{2}{3},\quad\sigma_{1B}=\frac{1}{3},\\
		\sigma_{2A}=&\frac{1+\alpha}{3},\quad\sigma_{2R}=\frac{1-2\alpha}{3},\quad\sigma_{2B}=\frac{\alpha-2}{3}, \\
		\alpha=&\frac{\chi_{AB}+\chi_{AR}-\chi_{BR}}{2\chi_{AB}}.
		\nonumber
	\end{split}
\end{equation}

The partition function of single chain $Q$ is determined by
\begin{equation}
	Q=\frac{1}{V}\int q_{B}(\br,s)q^{\dagger}_{B}(\br,s)~d\br, \quad s\in I_{B},
	\label{eq:Q}
\end{equation}
where flexible forward propagator $q_{B}(\br,s)$ describes the 
probability of finding the $s$-th segment at a spatial position 
$\br$  ranging from $s=0$ to $s=f_{B}$ under the mean field $\omega_{B}(\br)$.
Similarly, the flexible backward propagator $q_{B}^{\dagger}(\br,s)$ represents the 
probability from $s=f_{B}$ to $s=0$. 
Both the $q_{B}(\br,s)$ and $q_{B}^{\dagger}(\br,s)$ satisfy the modified diffusion equations (MDEs)
\begin{equation}
	\begin{split}
		\frac{\partial}{\partial s}q_{B}(\br,s)=&\nabla_{\br}^2q_{B}(\br,s)-\omega_{B}(\br)q_{B}(\br,s),\\
		q_{B}(\br,0)=&1,\quad s\in I_{B},
	\end{split}
	\label{model:coilPDEB}
\end{equation}
\begin{equation}
	\begin{split}
		\frac{\partial}{\partial s}q_{B}^\dagger(\br,s)=&\nabla_{\br}^2q_{B}^\dagger(\br,s)-\omega_{B}(\br)q_{B}^\dagger(\br,s),\\
		q_{B}^\dagger(\br,0)=&\int q_{R_1}(\br,\bu,f_{R_{1}})q_{R_2}(\br,\bu,f_{R_{2}})~d\bu,\quad s\in I_{B}.\\
	\end{split}
	\label{model:coilPDEplusB}
\end{equation}

The propagators $q_{A_{1}}(\br, s)$, $q_{A_{2}}(\br,s)$,  
$q^{\dagger}_{A_{1}}(\br, s)$, and $q^{\dagger}_{A_{2}}(\br,s)$
of flexible $A_{1}$ and $A_{2}$ blocks satisfy similar MDEs
\begin{equation}
	\begin{split}
		\frac{\partial}{\partial s}q_{A_{1}}(\br,s)&=\varepsilon^2\nabla_{\br}^2q_{A_1}(\br,s)-\omega_{A}(\br)q_{A_1}(\br,s),\\
		q_{A_{1}}(\br,0)&=1,\quad  s \in I_{A_{1}},
	\end{split}
	\label{model:coilPDEA1}
\end{equation}

\begin{equation}
	\begin{split}
		\frac{\partial}{\partial s}q_{A_{1}}^\dagger(\br,s)=&\varepsilon^2\nabla_{\br}^2q_{A_{1}}^\dagger(\br,s)-\omega_{A}(\br)q_{A_{1}}^\dagger(\br,s),\\
		q_{A_{1}}^\dagger(\br,0)=&\int q_{R_1}^\dagger(\br,\bu,f_{R_{1}})~d\bu,\quad  s \in I_{A_{1}},
	\end{split}
	\label{model:coilPDEplusA1}
\end{equation}

\begin{equation}
	\begin{split}
		\frac{\partial}{\partial s}q_{A_{2}}(\br,s)&=\varepsilon^2\nabla_{\br}^2q_{A_2}(\br,s)-\omega_{A}(\br)q_{A_2}(\br,s),\\
		q_{A_{2}}(\br,0)&=1,\quad  s \in I_{A_{2}},
	\end{split}
	\label{model:coilPDEA2}
\end{equation}

\begin{equation}
	\begin{split}
		\frac{\partial}{\partial s}q_{A_{2}}^\dagger(\br,s)=&\varepsilon^2\nabla_{\br}^2q_{A_{2}}^\dagger(\br,s)-\omega_{A}(\br)q_{A_{2}}^\dagger(\br,s),\\
		q_{A_{2}}^\dagger(\br,0)=&\int q_{R_2}^\dagger(\br,\bu,f_{R_{2}})~d\bu,\quad  s \in I_{A_{2}},
	\end{split}
	\label{model:coilPDEplusA2}
\end{equation}
where $\varepsilon=b_{A}/b_{B}$ measures the conformational asymmetric ratio of components $A$ and $B$ statistical segment lengths.

The forward propagator $q_{R_{j}}(\br,\bu,s)$ ($j = 1,\,2$) of the semi-flexible block, physically represents the probability of finding the $s$-th segment, from the endpoint $s=f_{A_j}$ to $s=f_{A_j}+f_{R_{j}}$ at spatial position $\br$ with orientation  $\bu$ under the mean field $\omega_{R}$. These propagators satisfy the ``convection diffusion'' equations

\begin{equation}
	\begin{split}
		\frac{\partial}{\partial s}q_{R_1}(\br,\bu,s)=&-\beta\bu\cdot\nabla_{\br}q_{R_1}(\br,\bu,s)\\
		-\Gamma(\br,\bu)&q_{R_1}(\br,\bu,s)+\frac{1}{2\lambda}\nabla_{\bu}^2q_{R_1}(\br,\bu,s),\\
		q_{R_1}(\br,\bu,0)=&\frac{q_{A_1}(\br,f_{A_{1}})}{2\pi}, \quad  s\in I_{R_{1}}, 
	\end{split}
	\label{model:rodPDER1}
\end{equation}
\begin{equation}
	\begin{split}
		\frac{\partial}{\partial s}q_{R_2}(\br,\bu,s)=&\beta\bu\cdot\nabla_{\br}q_{R_2}(\br,\bu,s)\\
		-\Gamma(\br,\bu)&q_{R_2}(\br,\bu,s)+\frac{1}{2\lambda}\nabla_{\bu}^2q_{R_2}(\br,\bu,s),\\
		q_{R_2}(\br,\bu,0)=&\frac{q_{A_2}(\br,f_{A_{2}})}{2\pi}, \quad  s\in I_{R_{2}}, 
	\end{split}
	\label{model:rodPDER2}
\end{equation}
where $\Gamma(\br,\bu)=\omega_{R}(\br)-\bM(\br):(\bu\bu-\frac{1}{2}\bI)$ is $\br$, $\bu$ dependent field,
$\beta=(b_{R}/b_{B})(6N)^{1/2}$ is the aspect ratio of the rods.
Similarly, backward propagators of the semi-flexible blocks $R_{1}$ and $R_{2}$ can be written as
\begin{equation}
	\begin{split}
		\frac{\partial}{\partial s}q_{R_1}^\dagger(\br,\bu,s)=&\beta\bu\cdot\nabla_{\br}q_{R_1}^\dagger(\br,\bu,s)\\
		-\Gamma(\br,\bu)&q_{R_1}^\dagger(\br,\bu,s)
		+\frac{1}{2\lambda}\nabla_{\bu}^2q_{R_1}^\dagger(\br,\bu,s),\\
		 q^{\dagger}_{R_{1}}(\br,\bu,0) = 
			&\frac{q_{B}(\br,f_{B})q_{R_{2}}(\br,\bu,f_{R_{2}})}{2\pi}, \quad  s\in I_{R_{1}},
	\end{split}
	\label{model:rodPDEplusR1}
\end{equation}

\begin{equation}
	\begin{split}
		\frac{\partial}{\partial s}q_{R_2}^\dagger(\br,\bu,s)=&-\beta\bu\cdot\nabla_{\br}q_{R_2}^\dagger(\br,\bu,s)\\
		-\Gamma(\br,\bu)&q_{R_2}^\dagger(\br,\bu,s)
		+\frac{1}{2\lambda}\nabla_{\bu}^2q_{R_2}^\dagger(\br,\bu,s),\\
		q^{\dagger}_{R_{2}}(\br,\bu,0) = 
		&\frac{q_{B}(\br,f_{B})q_{R_{1}}(\br,\bu,f_{R_{1}})}{2\pi},\quad  s\in I_{R_{2}}.
	\end{split}
	\label{model:rodPDEplusR2}
\end{equation}

The SCFT equations obtained from the first-order variational derivative of the free energy with respect to the field function are
\begin{equation}
	\begin{split}
		&\phi_{A}(\br)+\phi_{B}(\br)+\phi_{R}(\br)-1=0,\\
		&\frac{1}{2N\zeta_{1}}\mu_{1}(\br)-\sigma_{1A}\phi_{A}(\br)-\sigma_{1R}\phi_{R}(\br)-\sigma_{1B}\phi_{B}(\br)=0,\\
		&\frac{1}{2N\zeta_{2}}\mu_{2}(\br)-\sigma_{2A}\phi_{A}(\br)-\sigma_{2R}\phi_{R}(\br)-\sigma_{2B}\phi_{B}(\br)=0,\\
		&\frac{1}{\eta N}\bM(\br)-\bS(\br)=0,
		\label{eq:mu}
	\end{split}
\end{equation}

\begin{equation}
	\begin{split}
		\phi_{A}(\br)=&\frac{1}{Q}\bigg(\int_{I_{A_1}}q_{A_{1}}(\br,s)q^{\dagger}_{A_{1}}(\br,s)~ds\\
		&+\int_{I_{A_2}}q_{A_{2}}(\br,s) q^{\dagger}_{A_{2}}(\br,s)~ds  \bigg),
	\end{split}
\end{equation}

\begin{equation}
	\phi_{B}(\br)=\frac{1}{Q}\int_{I_{B}} q_{B}(\br,s)q_{B}^\dagger(\br,s)~ds,
\end{equation}

\begin{equation}
	\begin{split}
		\phi_{R}(\br)=&\frac{2\pi}{Q}\bigg(\int_{ I_{R_{1}}}\int 
		q_{R_1}(\br,\bu,s)q_{R_1}^\dagger(\br,\bu,s)~d\bu\,ds\\
		&+\int_{ I_{R_{2}}}\int q_{R_2}(\br,\bu,s)q_{R_2}^\dagger(\br,\bu,s)~d\bu\,ds\bigg),
	\end{split}
\end{equation}

\begin{equation}
	\begin{split}
		\bS(\br)=&\frac{2\pi}{Q}\bigg(\int_{I_{R_1}}\int q_{R_1}(\br,\bu,s)\big(\bu\bu-\frac{1}{2}\bI\big)q_{R_1}^\dagger(\br,\bu,s)~d\bu\,ds\\
		&+\int_{I_{R_{2}}}\int q_{R_2}(\br,\bu,s)\big(\bu\bu-\frac{1}{2}\bI\big)q_{R_2}^\dagger(\br,\bu,s)~d\bu\,ds\bigg),
	\end{split}
\end{equation}
where $\phi_{\alpha}(\br)$ ($\alpha\in \{A,\,B,\,R\}$) and $\bS(\br)$ are the monomer density of the $\alpha$-block and
the orientational order parameter, respectively.

Theoretical study of the phase behaviour of complex block copolymer systems within the SCFT framework generally follows two steps\,\cite{xu2013strategy,jiang2013analytic, jiang2015self}. The first step is to construct a library of candidate structures, which should contain as many candidate phases as possible. The construction of the candidate phases is inspired by relevant experimental and simulated findings, as well as theoretical considerations\,\cite{xu2013strategy, jiang2015self}. The second step uses an accurate and efficient algorithm to calculate the free energies of these candidate phases and then analyzes their relative stability. 
The phase diagram is then constructed by comparing the free energies of all candidate phases.

In the current study, we are interested in the stability of polygonal phases and their transition sequences. These two-dimensional polygonal phases can be regarded as columnar structures because of their homogeneity perpendicular to the polygonal plane. 
For these two-dimensional phases, the computations can be confined to two-dimensional space. The orientational calculation can be realized on the unit circle.
The most time-consuming step of solving the SCFT equations is computing these propagators, which are solutions of partial different equations. 
We employed the fourth-order backward differentiation\,\cite{cochran2006stability} and fourth-order Runge-Kutta methods\,\cite{butcher2003numerical} to solve the flexible and semi-flexible propagators equation, respectively.
The pseudo-sepectral method is used to treat both spatial and orientational variables due to periodic boundary conditions\,\cite{tzeremes2002efficient, rasmussen2002improved,jiang2010spectral}. 
An accelerated hybrid scheme that combines alternate iteration and conjugate gradient methods is utilized to 
search for the equilibrium states\,\cite{liang2015efficient} and optimize the computational box.
We carry out a parallel implementation in C++ language, utilizing the FFTW-MPI package\,\cite{frigo2005design}, to accelerate the SCFT computation.
Sec.\,\ref{SI:methods} in the Supporting Information (SI) presents a detailed description of these algorithms.

\section{Results and discussion}
\label{sec:results}
\begin{figure*}[!htbp]
	\centering
	\includegraphics[width=17cm]{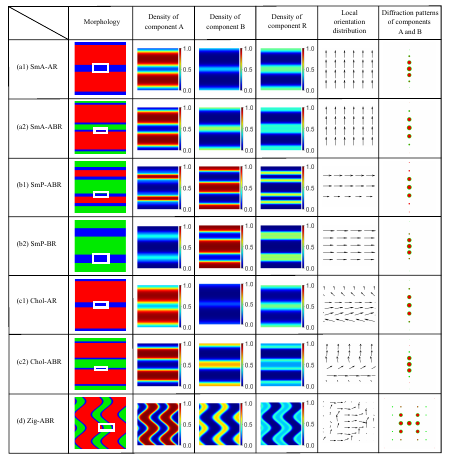}
	\caption{Layered structures self-assembled  from TLCMs. 
		$\rm \bf{Smectic}$-$\rm \bf{A}$ $\bf{phases}$ 
		(a1) SmA-AR; 
		(a2) SmA-ABR;
		$\rm \bf{Smectic}$-$\rm \bf{P}$ $\rm \bf{phases}$
		(b1) SmP-ABR; 
		(b2) SmP-BR;
		$\rm \bf{Cholesteric}$ $\rm \bf{phases}$ (c1) Chol-AR; (c2) Chol-ABR; and
		$\rm \bf{Zigzag}$ $\rm \bf{phase}$ (d) Zig-ABR.
		In the second column, red, green, and blue represent components $A$, $B$, and $R$ with high concentration, respectively.
		The third, fourth and fifth columns present the density distributions of components $A$, $B$ and $R$, respectively.
		The sixth column exhibits the orientation distribution 
		of the region framed by the white line in the second column.
		The last column shows the main diffraction peaks of components $A$ (red) and $B$ (green). 
	}
	\label{fig:L}
\end{figure*}

\textit{\textbf{Equilibrium phases.}} 
Based on extensive simulations, we obtained ten layered and sixteen polygonal phases as candidate phases for the TLCMs. 
The layered phases include smectic-A (SmA-AR, SmA-ABR, SmA-AB), smectic-P (SmP-ABR, SmP-BR, SmP-AB), cholesteric (Chol-AR, Chol-ABR, Chol-AB), and zigzag (Zig-ABR) phases. More detail of smectic phases can refer to SI, Sec.\,\ref{SI:results}. 
The density distributions of components $A$, $B$, $R$, and local orientation distribution of component $R$ in these layered phases are presented in Fig.\,\ref{fig:L} and SI, Fig.\,\ref{fig:LAB}.
The diffraction patterns obtained by Fourier transformation (see SI,\, \ref{subsec:FFT} for details) of the density distributions are also presented in these figures. The primary diffraction patterns of components $A$ and $B$ are marked with red and green dots, respectively. The sizes of these dots are proportional to the intensities of diffraction peaks. We scale the size of diffraction peak dots of component $A$ to be smaller than those of $B$ to ensure that the main green dots will not be obscured by the red dots.

\begin{figure*}[!hbpt]
	\centering
	\includegraphics[width=17cm]{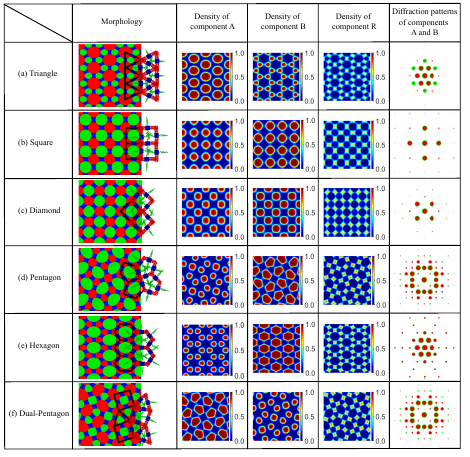}
	\caption{Simple polygonal phases self-assembled from TLCMs. 
	  (a) Triangle; (b) Square; (c) Diamond;
		(d) Pentagon; (e) Hexagon;
		(f) Dual-Pentagon. 
		The second column presents the morphologies combined with
		schematic arrangement diagrams, in which $A$-, $B$-, and $R$-rich domains are plotted in red, green and blue colors, respectively.
		The third, fourth and  fifth columns show the density distributions of components $A$, $B$ and $R$, respectively. The last column shows the main diffraction peaks of components $A$ (red) and $B$ (green).  
	}
	\label{fig:simplehoney}
\end{figure*}

\begin{figure*}[!hbpt]
	\centering
	\includegraphics[width=16cm]{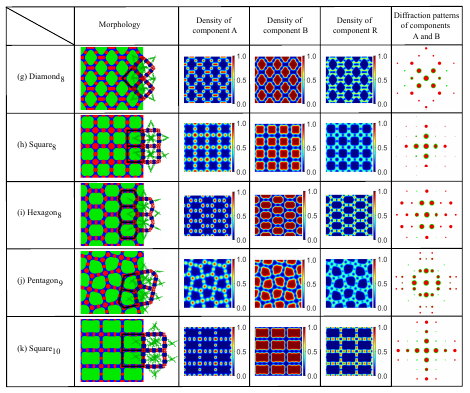}
	\caption{Giant polygonal phases: 
		(g) $\rm Diamond_{8}$; (h) $\rm Square_{8}$; 
		(i) $ \rm Hexagon_{8}$; (j) $\rm Pentagon_{9}$; 
		(k) $\rm Square_{10}$. 
		The meanings represented by subfigures are similar with Fig.\,\ref{fig:simplehoney}.
	}
	\label{fig:gainthoney1}
\end{figure*}

\begin{figure*}[!hbpt]
	\centering
	\includegraphics[width=16cm]{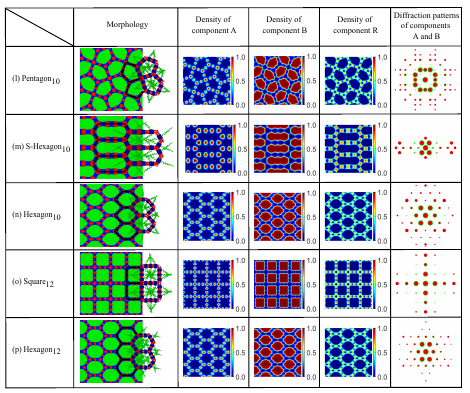}
	\caption{Giant polygonal phases: 
		 (l) $\rm Pentagon_{10}$; (m) S-$\rm Hexagon_{10}$;
		(n) $\rm Hexagon_{10}$; (o) $\rm Square_{12}$; 
		(p) $\rm Hexagon_{12}$. 
The meanings of each subfigure are similar with Fig.\,\ref{fig:simplehoney}.
	}
	\label{fig:gainthoney2}
\end{figure*}

The polygonal phases are classified into simple polygons (Fig.\,\ref{fig:simplehoney}) and giant polygons (Fig.\,\ref{fig:gainthoney1} and Fig.\,\ref{fig:gainthoney2}), based on the number of $R$-rich domains on the polygonal edges. 
In the simple polygons,
the number of polygonal edges is equal to the number of $R$-rich domains, whereas in the giant polygons, the number of polygonal edges is smaller than the number of $R$-rich domains.
The naming rules for these polygons are determined by both their polygonal shape (PS) and the number of $R$-rich domains (NR) on the polygonal edges, denoted as $\rm {PS}_{NR}$.  
For the simple polygons, the subscript is omitted. Figs.\,\ref{fig:simplehoney}-\ref{fig:gainthoney2} display the polygonal structures combined with molecular arrangement diagrams, the density distributions of components $A$, $B$, and $R$, and diffraction patterns of components $A$ and $B$.
In the simple polygons, the edges, vertices, and interiors of the Triangle, Diamond, Square, Pentagon, and Hexagon are composed of $R$-, $A$-, and $B$-rich domains.
The density distribution of $A$- and $B$-rich domains in the Dual-Pentagon\,\cite{chen2005liquid} exhibits a reciprocal relationship to that of the Pentagon, as illustrated in Fig.\,\ref{fig:simplehoney}\,(f).
The phases in Fig.\,\ref{fig:gainthoney2}\,(m) and (n) are named as $\rm Hexagon_{10}$ having a hexagonal shape containing ten $R$-rich domains.
The phase depicted in Fig.\,\ref{fig:gainthoney2}\,(m) has more pronounced stretching on the $B$-rich domains, 
causing  deformation of the hexagonal shape. 
This phase is named S-$\rm Hexagon_{10}$, with the `S' prefix indicating more stretching on the $B$-rich domains.

We focus on studying the influence of the side chain length and the block-block interactions on the stability of polygonal phases. To ensure the stability of polygonal phases, 
a specific set of parameters,
$\varepsilon=1$, $\lambda =300$, $\beta=6$,  $\eta=0.35$,  $f_{R_{1}}=f_{R_{2}}=0.10$,  
$f_{A_{1}}=(1-f_{B}-f_{R_{1}}-f_{R_{2}})/2$ and $f_{A_{2}}=f_{A_{1}}$, 
are selected, while the rest of parameters could vary.
To guarantee sufficient precision of the SCFT calculations,
we scan the phase space by primarily using discrete grids according to ordered phases and interaction strengths (see SI, Tab.\,\ref{tab:grid}). The termination criterion of self-consistent field iteration is the free energy difference between two consecutive iterations less than $10^{-8}$.
For convenience, we designate $\chi_{AB}$ as $\chi$ and use it as a reference, and express $\chi_{BR}$ and $\chi_{AR}$ as functions of $\chi$.

\begin{figure*}[!htbp]
	\centering
	\includegraphics[width=14cm]{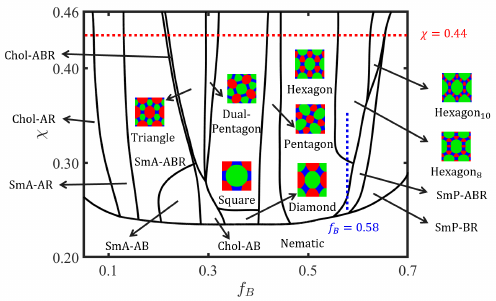}
	\caption{Phase diagram of TLCMs in the $\chi$-$f_{B}$ plane with $\chi_{AR}=\chi+0.04$, 
		$\chi_{BR}=\chi-0.02$, $N=100$, $f_{R_{1}}=f_{R_{2}}=0.10$. 
		The red and blue dashed lines mark the phase sequence of $\chi=0.44$ and $f_{B}=0.58$, respectively. }
	\label{fig:boundary}
\end{figure*}

\textit{\textbf{Stability of polygonal phases.}}
In the experiments, the end $A$ blocks can form hydrogen bonds\,\cite{tschierske2007liquid, tschierske2013development}. We can use
attractive $A$-$A$ interactions $\bar{\chi}_{AA}$ with negative value to model the hydrogen-bond interaction, arising the variation of effective Flory-Huggins parameters $\chi_{ij}$\,\cite{lee2006hydrogen}. 
Consequently, we investigate the influence of 
the attractive $A$-$A$ interactions of stabilizing polygonal phases. 
We simulate the phase behaviour for distinct $\chi_{ij}$ with $\chi\in [0.36,0.44]$. 
Seven phase diagrams in the $(\chi,f_B)$-plane have been constructed  with combinations of the parameters $\chi_{ij}$, as presented in SI, Fig.\,\ref{fig:seven}. 
The free energy difference of determining the phase boundaries is about $10^{-4}$.
It is evident that seven phase diagrams exhibit similar phase behaviours. 
As $f_{B}$ increases, phase transitions occur, from layered structures, to simple polygons, to giant polygons, and then to layered structures again. 
These results demonstrate that a slight perturbation of $\chi_{ij}$ has negligible influence on the relatively stability of candidate structures, and only leads a slight change in phase boundaries.
These phase diagrams also allow us to 
systematically investigate the impact of interaction strength by considering only one of the seven cases. Specially, we expand $\chi$ to $[0.20, 0.46]$, with $\chi_{AR}=\chi+0.04,\, \chi_{BR}=\chi-0.02$, and vary $f_B$.  
A detailed phase diagram, as shown in Fig.\,\ref{fig:boundary}, can be constructed with respect to $f_{B}$ and $\chi$.
This phase diagram presents much rich phase behaviours by varying $\chi$ and $f_B$. 
In the following, we will carefully analyze the impact of parameters $\chi$ and $f_{B}$ on the phase transitions.

\begin{figure}[h]
	\centering
	\includegraphics[width=14cm]{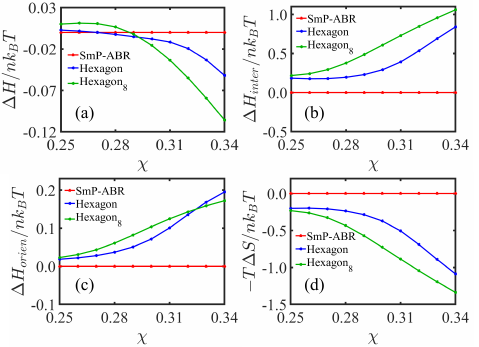}
	\caption{(a) Free energy, (b) interfacial energy, (c) orientation interaction energy and (d) entropic
 energy of the $\rm Hexagon$ (blue line) and $\rm Hexagon_{8}$ (green line) relative to the SmP-ABR (red line) along
increasing $\chi$ values for fixed $\chi_{AB}=\chi,\,\chi_{AR}=\chi+0.04,\,\chi_{BR}=\chi-0.02$, $f_{B}=0.58$, $f_{R_{1}}=f_{R_{2}}=0.10$, and $N=100$.
}
	\label{fig:LHH}
\end{figure}

\textit{-- Influence of interaction parameter $\chi$ --} 
We examine the phase transition path from layers to simple polygons, and to giant polygons by varying $\chi$ and fixing $f_B$.  
When $f_B=0.58$, a phase sequence emerges, taking the system from the SmP-ABR ($0.25\leq \chi < 0.27$) to the simple Hexagon polygon $(0.27\leq \chi < 0.293)$, and to the giant $\rm Hexagon_{8}$ polygon ($\chi \ge 0.293$), as shown by the blue dash line in Fig.\ref{fig:boundary}. 
The energy curves, taking SmP-ABR phase as the baseline, are plotted in Fig.\,\ref{fig:LHH}\,(a). To better analyze the factors of influencing the stability, we split the free energy into three parts, the interfacial energy $H_{inter}/nk_{B}T$, the orientation interaction energy $H_{orien}/nk_{B}T$, and the entropy energy $- T S/nk_{B}T$, see the definition in Eq.\,\eqref{eq:energySplit}. 
The density distribution demonstrates that polygonal structures have more $A$-, $B$- and $R$-rich subdomains than the layered SmP-ABR, and the $\rm Hexagon_{8}$ has the most subdomains among three patterns.
As shown in Fig.\,\ref{fig:LHH}\,(b), 
the more subdomains the structure has, the larger interfacial energy the system has.  
On the other hand, more subdomains provide an opportunity that molecular chains have much freedom of stretch, thus leading to a lower entropy energy, as Fig.\,\ref{fig:LHH}\,(d) illustrates. 
Meanwhile, the orientation distribution of polygonal phases, including the $\rm Hexagon$ and the $\rm Hexagon_{8}$, are more disordered than that of the layered SmP-ABR (see Fig.\,\ref{figapp:orienL-T}), arising a larger orientation interaction energy (see Fig.\,\ref{fig:LHH}\,(c)).
Interestingly, for the polygonal phases, as $\chi$ increases, the orientation interaction energies of the $\rm Hexagon$ and the $\rm Hexagon_{8}$ have a intersection point at $\chi =0.324$. Below this intersection, the Hexagon has lower orientation interaction energy. Otherwise, the $\rm Hexagon_{8}$
has lower value. The reason can be attribute to the Flory-Huggins interaction. 

\begin{figure}[!htbp]
	\centering
	\includegraphics[width=14cm]{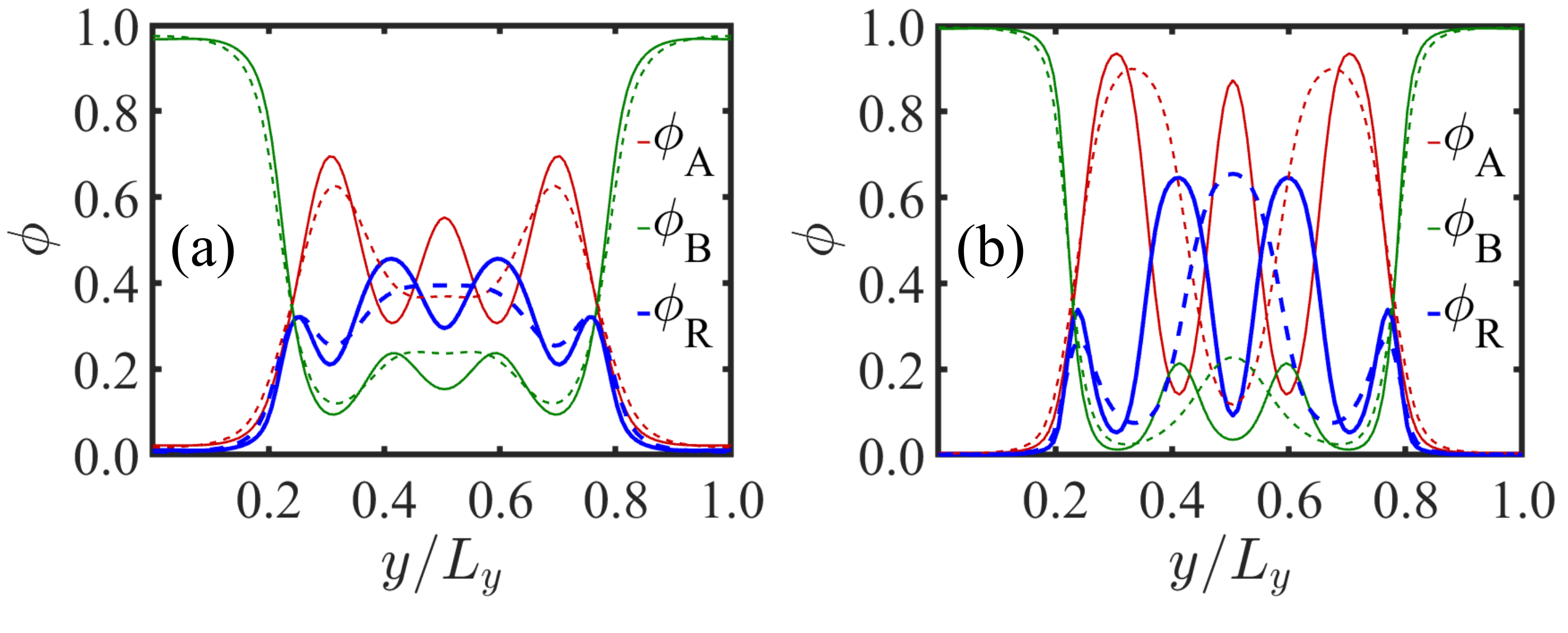}
	\caption{The solid and the dashed lines represent the density distributions of the $\rm Hexagon_{8}$ and $\rm Hexagon$, respectively. 
	(a) $\chi=0.28$, (b) $\chi=0.34$ for fixed $L_{x}/4$ (see SI, Fig.\,\ref{figapp:Hex}), $\chi_{AB}=\chi,\,\chi_{AR}=\chi+0.04,\,\chi_{BR}=\chi-0.02$, $f_{B}=0.58$, $f_{R_{1}}=f_{R_{2}}=0.10$, and $N=100$.
	}
	\label{fig:distrubution88}
\end{figure}

As shown in Fig.\,\ref{fig:distrubution88}, 
with an increase of $\chi$ from 0.28 to 0.34, the peak of $R$-rich domain in the Hexagon increases from 0.4 to 0.63, with an increment of 0.23. While the peak of $R$-rich domain in the $\rm Hexagon_{8}$ rises from 0.47 to 0.65, with an increment of 0.18.
This indicates that as $\chi$ increases, the $A$-$R$ and $B$-$R$ repulsion interactions in the Hexagon increase faster than that of $\rm Hexagon_{8}$, yielding a disordered orientation distribution. Consequently, the orientation interaction energy in the Hexagon phase gradually exceeds that in the $\rm Hexagon_{8}$ phase.
During the subtle competition among three parts of energies, the above-mentioned phase sequence emerges.

\begin{figure*}[!hbpt]
	\centering
	\includegraphics[width=14cm]{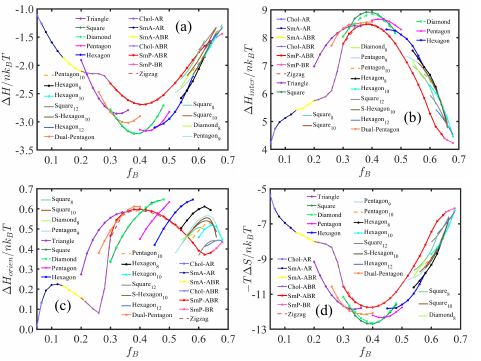}
	\caption{
	(a) Free energy, (b) interfacial energy, (c) orientation interaction energy, 
	and (d) entropic energy of the candidate phases
	relative to the homogeneous phase by varying $f_{B}$ when 
	$\chi_{AB}=0.44,\, \chi_{AR}=0.48,\, \chi_{BR}=0.42$, $N=100$, and $f_{R_{1}}=f_{R_{2}}=0.10$.}
	\label{fig:energy44}
\end{figure*}

\textit{-- Influence of volume fraction $f_{B}$ --}
Here we consider the effect of volume fraction $f_B$ on the stability of candidate patterns. For a fixed $\chi=0.44$, an interesting phase sequence of Chol-AR  $\rightarrow$ SmA-AR  $\rightarrow$ SmA-ABR  $\rightarrow$ Chol-ABR  $\rightarrow$ Triangle 
$\rightarrow$ Dual-Pentagon
$\rightarrow$ Square  $\rightarrow$ Pentagon  $\rightarrow$ Hexagon  $\rightarrow$ $\rm Hexagon_{8}$ $\rightarrow$ $\rm Hexagon_{10}$ $\rightarrow$ SmP-AB appears as $f_{B}$ increases. 
The free energy curves of these structures relative to the homogeneous phase are plotted in Fig.\,\ref{fig:energy44}\,(a). 
To better analyze the factors of influencing stability, we again separate the free energy into three parts, the interfacial energy, the orientational interaction energy, and the entropic energy. 
When the volume factor $f_B$ is smaller than $0.14$, the length of $B$ subchain is too small to separate from $R$-rich domain, leading to the formation of the two layered phases of Chol-AR ($0.045<f_B \leq 0.065$) and SmA-AR ($0.065<f_B \leq 0.14$).
As $f_B$ increases, monomer-$B$ can condensate into the new $B$-rich layer, causing three layered patterns of SmA-ABR ($0.14<f_B\leq 0.214$) and Chol-ABR ($0.214<f_B\leq 0.218$). It is noted that the SCFT calculation predicts stable Chol-AR and Chol-ABR, which were not observed experimentally. This discrepancy might be attributed to the thermodynamic fluctuations, which are not accurately captured by the SCFT.
When $0.218<f_B \leq 0.657$, the system enters the stability region of polygonal phases. Compared to layered phases, polygonal structures possess more $A$-, $B$-, and $R$-rich subdomains which increase the interfacial energy (see Fig.\,\ref{fig:energy44}\,(b)). 
These divided subdomains also disrupt the parallel arrangement of rigid blocks resulting in a small increase  of the orientation interaction energy (see Fig.\,\ref{fig:energy44}\,(c)).
On the other hand, these subdomains in polygonal phases also make the chain arrangement more flexible, thus increasing the  configurational entropy (see Fig.\,\ref{fig:energy44}\,(d)). The arising entropy energy is more than the unfavorable interfacial and orientation energies, driving the phase transition from layered phases to polygonal phases. In the range of polygonal phases, as the relative length of $B$ block increases, the interior of $B$-rich domain swells. To alleviate the packing frustration of compressing $B$-rich domain, the system increases the number of polygonal edges, driving a phase transition from simple polygons to giant polygons when $f_B>0.56$.
Meanwhile, several novel metastable giant polygons are also observed, including $\rm Square_{8}$, $\rm Square_{10}$, $\rm Square_{12}$, $\rm Diamond_{8}$, $\rm Pentagon_{9}$, and $\rm Hexagon_{12}$ which might be stable at more strong segregation.

\begin{figure*}[!htbp]
	\centering
	\includegraphics[width=13cm]{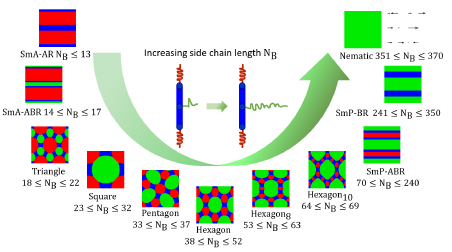}
	\caption{ Phase transition as $N_{B}$ increases with $\chi_{AB}=0.36$, $\chi_{AR}=0.40$,
		$\chi_{BR}=0.34$, 
		$N_{A_{1}}=N_{A_{2}}=10$, and $N_{R_{1}}=N_{R_{2}}=10$. }
	\label{fig:sidechain}
\end{figure*}
\textit{-- Influence of the side chain length --} 
The above simulations have examined the influence of Flory-Huggins interaction $\chi$ and relative volume factor $f_B$ on the stability of polygonal phases. Previous experiments have demonstrated that the TLCMs exhibit an interesting phase transition sequence of
SmA-AR $\rightarrow$ SmA$^{+}$ 
$\rightarrow$ Triangle $\rightarrow$ Diamond 
$\rightarrow$ Square $\rightarrow$  
Pentagon $\rightarrow$ Hexagon  
$\rightarrow$ $\rm Hexagon_{8}$ $\rightarrow$ 
$\rm Hexagon_{10}$ $\rightarrow$  $\rm Pentagon_{10}$
$\rightarrow$ Lamellar $\rightarrow$ bicontinuous 
cubic phases when the side chain length of TLCMs is increased\,\cite{cheng2003calamitic, tschierske2007liquid, tschierske2013development}.
In what follows we theoretically investigate this phase transition sequence by varying the side chain length.
To model the experimental systems, we use the monomers number $N_{i}$ ($i\in \{A_{1},\, A_{2},\, B,\, R_{1},\, R_{2}\}$) to describe the block length. Varying $N_i$ is equivalent to changing the length of the different blocks. Based on the aforementioned calculations, we fix a set of parameters $\chi_{AB} = 0.36$, $\chi_{AR} = 0.40$, $\chi_{BR} = 0.34$, $N_{A_{1}} = N_{A_{2}}= 10$, $N_{R_{1}}=N_{R_{2}}=20$, and change $N_B$.
With an increase of the side chain length $N_{B}$, 
the SCFT calculations obtain a phase transition sequence of 
SmA-AR $ \rightarrow $ SmA-ABR $ \rightarrow$ 
Triangle $\rightarrow$ Square 
$\rightarrow$ Pentagon $\rightarrow$ Hexagon $\rightarrow$ 
$\rm Hexagon_{8}$ $\rightarrow$ $\rm Hexagon_{10}$ $\rightarrow$ 
SmP-ABR $\rightarrow$ SmP-BR $\rightarrow$ Nematic phase, as shown in Fig.\,\ref{fig:sidechain}. 
The reason of forming the nematic phase when $351 \le N_B \le 370$ may be ascribed to the microphase separation of different chemical components. As a result, the concentration of a rigid backbone in the $R$-rich domain is high and can lead to the long-range orientation order.
The theoretical phase transition sequence is consistent with existing experimental observations\,\cite{cheng2003calamitic, tschierske2007liquid, tschierske2013development}. 
There are slight differences between the experimental observations and our theoretical predictions. 
For example, our theoretical results indicate that the $\rm Pentagon_{10}$ and Diamond phases are metastable (see SI, Fig.\,\ref{fig:N36}), while these phases were reported as stable ones in experiments. 
This discrepancy might be ascribed to the fact that our simulation parameters could be not entirely identical to the experimental conditions, or due to the use of Gaussian chain model to describe short flexible chains.

\section{Conclusion}
\label{sec:concluding}
In summary, we have established a SCFT model of TLCMs 
to investigate the formation and stability of polygonal phases.
The development of an accurate and efficient numerical method for SCFT equations enables us to construct a set of  phase diagrams by precisely computing the free energy of different self-assemble ordered structures. 
We examined the influence of the side chain length and the interaction strength on the stability of polygonal phases and their transitions.
We systematically analyzed the stability mechanism by examining the free energy.
The resulting phase transition sequences are in good agreement with experimental observation.
Several new metastable polygonal structures, and several smectic, cholesteric, and zigzag layers are also predicted in our study. 
These theoretical findings fill the gap
between theoretical understanding
and experimental observation
of the phases and phase transitions of TLCMs.
In the future, we will investigate complicated phases and phase transition in more liquid crystalline molecular systems based on the SCFT and advanced numerical methods developed in the current study.

\begin{suppinfo}
	\begin{itemize}
		\item  Free energy of homogeneous phase, Numerical methods, 
		Figs \ref{fig:schematic}-\ref{fig:N36}, 
		and Tab \ref{tab:grid} (PDF)
	\end{itemize}
\end{suppinfo}

\section{Acknowledgement}
The work is supported in part by the National Key R\&D Program of China (2023YFA1008802), the National Natural Science Foundation of China (12171412, 12288101), 
Postgraduate Scientific Research Innovation Project of Hunan Province (CX20220636), 
Postgraduate Scientific Research Innovation Project of Xiangtan University (XDCX2022Y061),
and the Natural Sciences and Engineering Research Council of Canada. We are also grateful to the High Performance Computing Platform of Xiangtan University for partial support of this work.

\bibliography{ref.bib}

\end{document}


%
%
%
%
%



\begin{figure}[H]
	\centering
	\includegraphics[width=15cm]{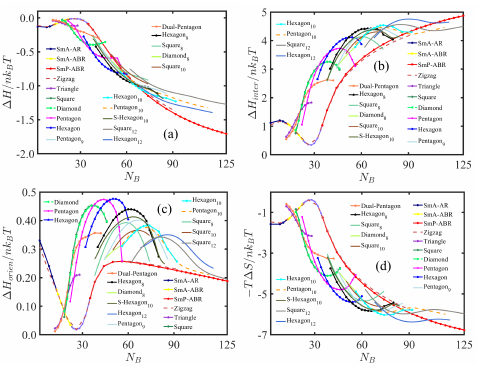}
	\caption{ (a) Free energy, (b) interfacial energy, 
	(c) orientation interaction energy, and (d) entropic energy of the candidate phases
	relative to the homogeneous phase along increasing $N_{B}$ values for fixed
	 $\chi_{AB}=0.36,\, \chi_{AR}=0.40,\, \chi_{BR}=0.34$, and $ N_{A_{1}}= N_{A_{2}}=N_{R_{1}}=N_{R_{2}}=10$.}
	\label{fig:N36}
\end{figure}

\setlength{\parindent}{2em}
\noindent
This PDF file includes:\\
\ref{SI:homogeneous} Free energy of homogeneous phase\\
\ref{SI:methods} Numerical methods\\
\ref{SI:results} Numerical results\\
\indent Figures \ref{tab:grid}-\ref{fig:N36}\\
\indent Table \ref{fig:schematic}

\newpage

\section{Free energy of homogeneous phase}
\label{SI:homogeneous}

In the T-shaped liquid crystalline molecules (TLCMs), the density and orientation distributions of the homogeneous phase are independent of spatial position, satisfying $\phi_{i}(\br)=f_{i},\, S(\br)=0$, where $i\in \{A,\,B,\,R\}$.
The expressions of field functions $\omega_{A}(\br)$, $\omega_{B}(\br)$, $\omega_{R}(\br)$ and 
$\mu_{+}(\br)$, $\mu_{1}(\br)$, $\mu_{2}(\br)$ are solely dependent on the model parameters, which are given by,
\begin{equation}
\begin{split}
\mu_{+}(\br) = & 0,\\
\mu_{1}(\br) = & 2N\zeta_{1}\bigg(\frac{1}{3}-f_{R_1}-f_{R_2}\bigg),\\
\mu_{2}(\br) = & 2N\zeta_{2}\bigg(\frac{\alpha-2}{3}+f_{A_{1}}+f_{A_{2}}+(1-\alpha)(f_{R_{1}}+f_{R_{2}})\bigg),\\
\omega_{A}(\br) = & \mu_{+}(\br) - \sigma_{1A}\mu_{1}(\br) -\sigma_{2A}\mu_{2}(\br),\\
\omega_{B}(\br) = & \mu_{+}(\br) - \sigma_{1B}\mu_{1}(\br) -\sigma_{2B}\mu_{2}(\br),\\
\omega_{R}(\br) = & \mu_{+}(\br) - \sigma_{1R}\mu_{1}(\br) -\sigma_{2R}\mu_{2}(\br).\\
\end{split}
\end{equation}
From the free energy expression in Eq.\,(\ref{H}), we also need to derive the analytical expression for the single-chain partition function $Q$. Based on the properties of the homogeneous phase, we can deduce the analytical expression for the propagators $q_{B}(\br,f_{B})$ and $q^{\dagger}_{B}(\br,0)$, which are independent of the spatial variable $\br$. Consequently, the analytical expression for the single-chain partition function $Q$ can be represented as, 
\begin{equation}
    Q=\frac{1}{2\pi}\exp{\{-\omega_{A}(f_{A_{1}}+f_{A_{2}}) -\omega_{R}(f_{R_{1}}+f_{R_{2}})-\omega_{B}(f_{B}) \}}.
\end{equation}
The analytical expression of the free energy of homogeneous phase is,
\begin{equation}
\begin{split}
H=&N\zeta_{1}\bigg(\frac{1}{3}-f_{R_{1}}-f_{R_{2}}\bigg)^2+N\zeta_{2}\bigg(\frac{\alpha-2}{3}+f_{A_{1}}+f_{A_{2}}+(1-\alpha)(f_{R_{1}}+f_{R_{2}})\bigg)^2\\
&-\log(\frac{1}{2\pi})-\bigg(-\omega_{A}(\br)(f_{A_{1}}+f_{A_{2}}) -\omega_{R}(\br)(f_{R_{1}}+f_{R_{2}})-\omega_{B}(\br)f_{B}\bigg).
\end{split}
\end{equation}

\section{Numerical methods}
\label{SI:methods}

Our calculations can be confined to a two-dimensional space with rectangular box $L_{x}\times L_{y}$.
The orientational calculation can be realized on the unit circle. 
We employ the Fourier pseudo-spectral method\,\cite{tzeremes2002efficient, rasmussen2002improved, jiang2010spectral} to handle spatial and orientational variables. 
The fourth-order backward differentiation\,\cite{cochran2006stability} and fourth-order 
Runge-Kutta methods\,\cite{butcher2003numerical} are employed
to solve the flexible and semi-flexible propagators equation 
for the chain contour variable, respectively. 
An accelerated hybrid scheme that combines alternate iteration and 
conjugate gradient methods is utilized to 
search for the equilibrium states\,\cite{liang2015efficient} and optimize the computational region.

\subsection{Fourier pseudo-spectral method}
\label{subsec:FFT}
The Fourier series expansion of the periodic function $q_{R_1}(\br, \bu, s)$ at discrete spatial positions $\br_{lj}$ and orientation positions $u_{m}$ is presented as follows,

\begin{align}
f(\br_{lj}, u_{m}, s)=\sum_{\bk \in \mathbb{K}, v\in \mathbb{V}}\hat{q}_{R_1}(\bk, v, s) e^{i (\calA
	\bk)^T \br_{lj}} e^{ivu_{m}},
\label{Fourier:expansion}
\end{align}
where $\calA$ is reciprocal lattice to the calculation lattice $\calB$, and the discrete Fourier coefficient $\hat{q}_{R_1}(\bk,v,s)$ can be calculated by using Fast
Fourier Transform(FFT). 
$\mathbb{K}$ and $\mathbb{V}$ are defined, 
\begin{align*}
\begin{split}
\mathbb{K}:=&\{\bk = (k_{x},\,k_{y}) \in \mathbb{Z}^2:\,
-N_{x}/2\leq k_{x}<N_{x}/2, -N_{y}/2\leq k_{y}<N_{y}/2\},\\
\mathbb{V}:=&\{v \in \mathbb{Z}:-N_{\theta}/2 \le v <N_{\theta}/2\}.
\end{split}
\end{align*}
where $N_{x}\times N_{y}$ and $N_{\theta}$ represent the discrete points of the spatial and directional variables.

A periodic function $\phi(\br)$ can be expanded as
\begin{align}
\phi(\br)=\sum_{\bk \in \mathbb{K}} \hat{\phi}(\bk) e^{i (\calA
	\bk)^T \br},
\end{align}
where the Fourier coefficients
\begin{align}
    \hat{\phi}(\bk) = \frac{1}{|\calB|} \int_{\calB} \phi(\br) e^{-i (\calA
	\bk)^T \br} \,d\br.
	\label{eq:Fcoeff}
\end{align}

\subsection{Fourth-order backward differentiation formula (BDF4)}
\label{subsec:BDF4}

The contour variable $s$ in the flexible propagators equation is discretized using the BDF4 method\,\cite{cochran2006stability}. Taking Eq.\,\eqref{model:coilPDEB} as an example, we express the transformation from $s_{n-1}$ to $s_{n}=s_{n-1}+\Delta s$ as follows,
\begin{align}
\begin{split}
&\frac{25}{12}q_{B}^{n}(\br)-4q_{B}^{n-1}(\br)+3q_{B}^{n-2}(\br)-\frac{4}{3}q_{B}^{n-3}(\br)+\frac{1}{4}q_{B}^{n-4}(\br)\\
&=\Delta s\left[\nabla^2
q_{B}(\br)^{n}-w(\br)\left(4q_{B}^{n-1}(\br)-6q_{B}^{n-2}(\br)\right.\right.\\
&\left.\left.+4q_{B}^{n-3}(\br)-q_{B}^{n-4}(\br)\right)\right].
\end{split}
\end{align}
In our calculations, the initial values for the first four steps of the BDF4 method are obtained using a special extrapolation method\,\cite{ranjan2008linear} based on the second-order operator-splitting scheme\,\cite{rasmussen2002improved}.

\subsection{Fourth-order Runge-Kutta (RK4) method}
\label{subsec:RK4}
The RK4 method\,\cite{butcher2003numerical} is used to discretize the contour variable $s$ of semi-flexible propagators. For example, Eq.\,\eqref{model:rodPDER1} can be rewritten as
\begin{align}
\frac{\partial}{\partial s}q_{R_{1}}(\br,\bu,s)=F(q_{R_{1}}(\br,\bu,s)),
\label{RK:ODE}
\end{align}
where $F(q_{R_{1}}(\br,\bu,s))=\beta \bu \cdot \nabla_{\br} q_{R_{1}}(\br,\bu,s)-
\Gamma(\br,\bu)q_{R_{1}}(\br,\bu,s)+\frac{1}{2\lambda}q_{R_{1}}(\br,\bu,s)$.
Definition:
\begin{equation}
\begin{aligned}
K_{1}&=F(q_{R_{1}}^{n-1}(\br,\bu)),
\\
K_{2}&=F(q_{R_{1}}^{n-1}(\br,\bu)+\Delta sK_{1}/2),
\\
K_{3}&=F(q_{R_{1}}^{n-1}(\br,\bu)+\Delta sK_{2}/2),
\\
K_{4}&=F(q_{R_{1}}^{n-1}(\br,\bu)+\Delta sK_{3}).
\end{aligned}
\end{equation}
From $s_{n-1}$ to $s_{n}=s_{n-1}+\Delta s$,
the RK4 method can be expressed as,
\begin{align}
q_{R_{1}}^{n}(\br,\bu)=q_{R_{1}}^{n-1}(\br,\bu)+\frac{\Delta s}{6}(K_{1}+2K_{2}+2K_{3}+K_{4}).
\end{align}

\subsection{Hybrid scheme (HS)}
\label{subsec:HS}

The HS minimizing a functional $H[\bu]$ is described as follows.
\begin{algorithm}[h]
	\caption{ Hybrid scheme}
	\label{alg:HS}
	\KwIn{initial value $\bu$, $k=0$.}
	$\bg_{0}=\frac{\delta H[\bu]}{\delta \bu}$,
	
	$\bs_{0}=-\bg_{0}$,
	
	\While {$\frac{\delta H[\bu]}{\delta \bu} \neq 0$}{
		$k = k+1$; 
		
		choose $\gamma_{k}$ to minimize $H[\bu_{k}+\gamma_{k}\bs_{k}]$,
		
		$\bu_{k+1}=\br_{k}+\gamma_{k}\bs_{k}$,
		
		$g_{k+1}=\frac{\delta H[\bu_{k+1}]}{\delta \bu}$,

		$\beta_{k+1}=\bigl(\bg^{T}_{k+1}\bg_{k+1}\bigr)/(\bg^{T}_{k}\bg_{k})$,
		
		$\bs_{k+1}=-\bg_{k+1}+\alpha_{c}\beta_{k+1}\bs_{k}$.
	}
	\label{algo}
\end{algorithm}
In this algorithm, $ \alpha_{c}$ serves as a hybrid factor used to modify the conjugate gradient direction.
When $\alpha_{c}=0$, the HS transforms into the alternative direction iteration method. Conversely, when $\alpha_{c}=1$, the HS converts to the conjugate gradient method. 
When $0 <\alpha_{c}< 1$, the algorithm combines the advantages of both the  alternative direction iteration and conjugate gradient methods.
The HS is capable of determining the maximum value of a function by adjusting the sign of the gradient.
Thus, we can use the HS to update the field functions $\mu_{+},\, \mu_{1},\, \mu_{2}$, and $\bM$, as well as the computational region $\mathcal{B}$, based on the orientation of the saddle point of SCFT. 
Specifically, $\partial H/\partial\mathcal{B}$ is calculated using the central difference method.

\section{Numerical results}
\label{SI:results}
\begin{table}[H]
	\centering
	\caption{Discretization grids.
	 $N_{x}\times N_{y}$ and $T$ are the discretization grid nodes for space and chain contour, respectively.
	 The number of orientation points is  $N_{\theta}=16$.
	 }
	\begin{tabular}{ccccc}
		\hline
		\multirow{2}*{}&
		\multicolumn{2}{c}{$0.20<\chi_{AB}<0.40$}&\multicolumn{2}{c}{$0.40\le\chi_{AB}\le 0.46$}\\
		\cmidrule(lr){2-5}
			  &  Layer phases & Polygons & Layer phases &    Polygons   \\
		\hline
		Nx   &61 & 81 &101 &121 \\
		Ny &61 & 81 &101&121 \\
		T &  200 & 200& 200&300 \\
		\hline
	\end{tabular}	
	\label{tab:grid}
\end{table}

\textit{\textbf{Layer-like phases}} 
Layered phases can be classified into different types  based on the angle $\theta$ between the layer normal $\bn$ and the orientational directions of the semi-flexible blocks\,\cite{liang2015efficient}.

\begin{figure}[H]
	\centering
	\subfigure[$\theta=0$.]{
		\includegraphics[width=4.5cm]{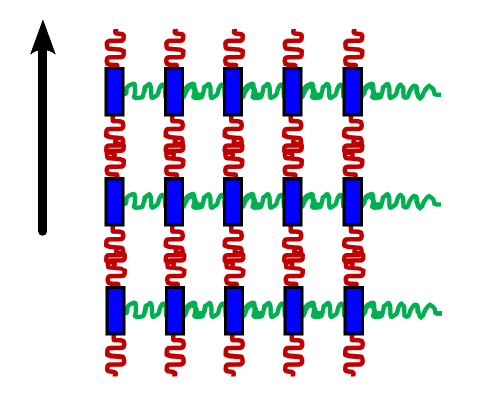}
	}
	\subfigure[$\theta= \pi/2$.]{
		\includegraphics[width=4cm]{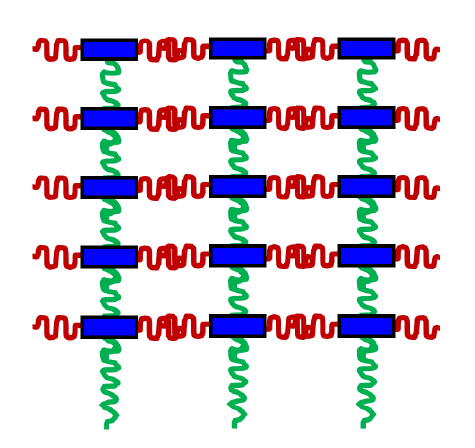}
	}
	\subfigure[$\theta$ changes regularly.]{
		\includegraphics[width=4cm]{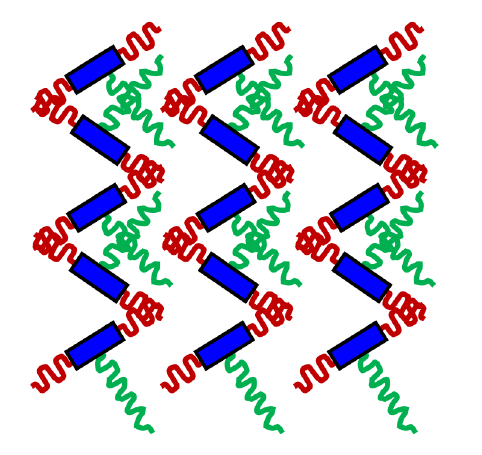}
	}
	\caption{The black arrow $n$ is the normal direction of layered phase, TLCMs schematic arrangement diagrams of (a) smectic-A phase, 
	(b) smectic-P phase, and (c) cholesteric phase.
	}
	\label{fig:schematic}
\end{figure}

($ \bm{\romannumeral1}$) If $\theta=0$, 
it is smectic-A (SmA) phase (TLCMs schematic arrangement diagrams see Fig.\,\ref{fig:schematic}\,(a)). 

($ \bm{\romannumeral2}$) If $\theta=\pi/2$, 
it is smectic-P (SmP) phase (TLCMs schematic arrangement diagrams see Fig.\,\ref{fig:schematic}\,(b)). 

($ \bm{\romannumeral3}$) 
If the orientations within a layer are arranged in parallel, 
and the orientations between adjacent layers continuously and uniformly rotate along the normal direction of the layers, 
the whole structure takes on a spiral shape. This is known as the cholesteric (Chol) phase (TLCMs schematic arrangement diagrams see Fig.\,\ref{fig:schematic}\,(c)).

\begin{figure}[H]
	\centering
		\includegraphics[width=16cm]{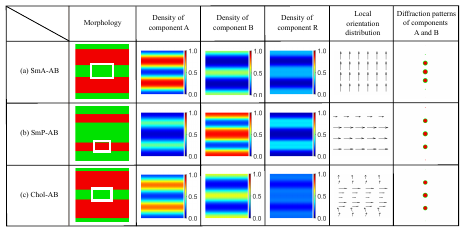}
		\caption{
		Layered structures self-assembled  from TLCMs. 
			(a) SmA-AB; 
		(b) SmP-AB;
		(c) Chol-AB.
		In the second column, red, green, and blue represent components $A$, $B$, and $R$ with high concentration, respectively.
		The third, fourth and fifth columns present the density distribution of  components $A$, $B$ and $R$, respectively.
		The sixth column exhibits the orientation distribution 
		of the region framed by the white line in the second column.
		The last column shows the main diffraction peaks of components $A$ (red) and $B$ (green). 
		}
	\label{fig:LAB}
\end{figure}

\begin{figure}[H]
	\centering	
	\subfigure[ $\chi_{AR}=\chi+0.04$, $\chi_{BR}=\chi-0.02$]{
		\includegraphics[width=7cm]{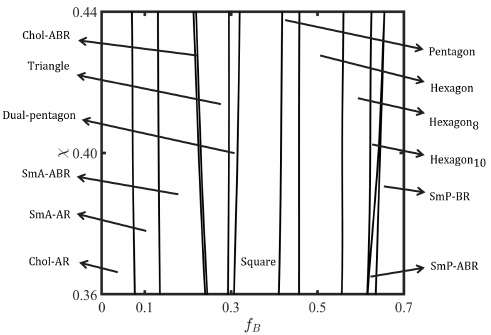}
	}
	\subfigure[ $\chi_{AR}=\chi-0.04$, $\chi_{BR}=\chi-0.02$]{
		\includegraphics[width=7cm]{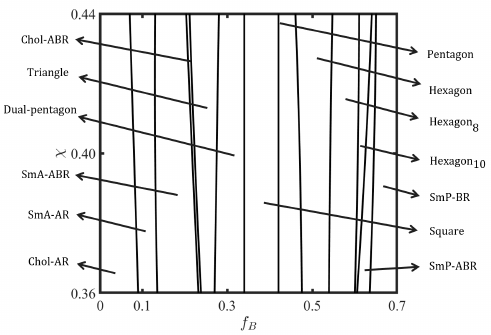}
	}
	\subfigure[ $\chi_{AR}=\chi-0.02$, $\chi_{BR}=\chi-0.04$,  ]{
		\includegraphics[width=7cm]{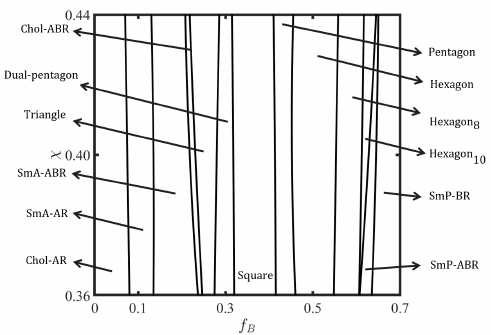}
	}
	\subfigure[ $\chi_{AR}=\chi-0.04$, $\chi_{BR}=\chi-0.04$]{
		\includegraphics[width=7cm]{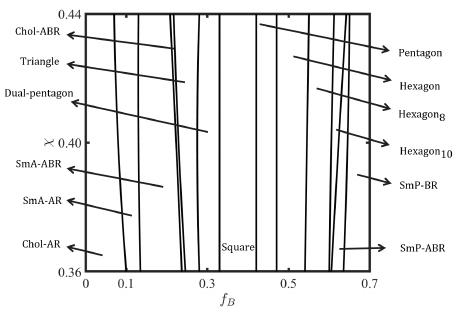}
	}
	\subfigure[ $\chi_{AR}=\chi+0.04$, $\chi_{BR}=\chi+0.04$]{
		\includegraphics[width=7cm]{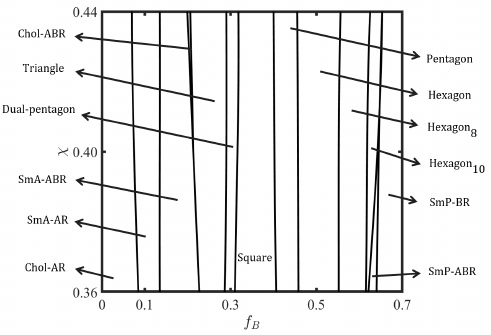}	
	}
	\subfigure[  $\chi_{AR}=\chi-0.04$, $\chi_{BR}=\chi$]{
		\includegraphics[width=7cm]{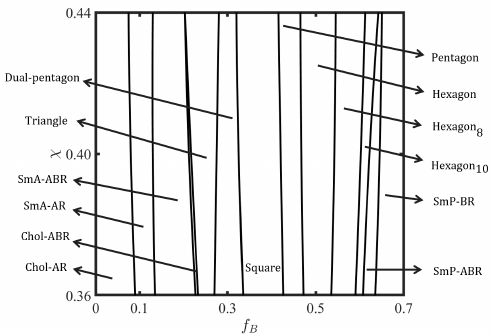}	
	}
	\subfigure[ $\chi_{AR}=\chi+0.04$, $\chi_{BR}=\chi$]{
		\includegraphics[width=7cm]{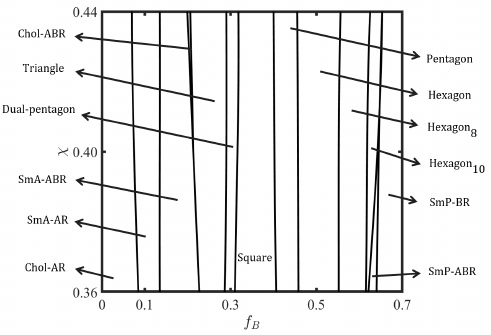}	
	}
	\caption{Seven phase diagrams in the $\chi-f_{B}$ plane of TLCMs with $N=100,\, f_{R_{1}}=f_{R_{2}}=0.10$.}
	\label{fig:seven}
\end{figure}

\begin{figure}[H]
    \centering
		\includegraphics[width=15cm]{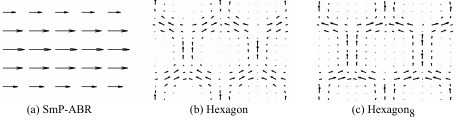}
	\caption{The orientation distribution of (a) SmP-ABR, (b) Hexagon and (c) $\rm Hexagon_{8}$ in the unit cell with
	$\chi_{AB}=0.34,\, \chi_{AR}=0.38,\, \chi_{BR}=032$, $N=100$, $f_{B}=0.58$ and $f_{R_{1}}=f_{R_{2}}=0.10$.}
	\label{figapp:orienL-T}
\end{figure}

\begin{figure}[H]
	\centering
	\includegraphics[width=12cm]{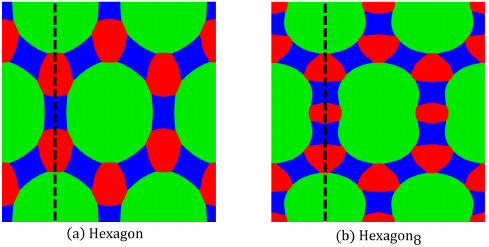}
	\caption{The density distribution of (a) Hexagon and (b) $\rm Hexagon_{8}$ in the unit cell at $L_{x}/4$ is marked by the black line with $\chi_{AB}=0.30,\, \chi_{AR}=0.34,\, \chi_{BR}=0 28$, $N=100$, $f_{B}=0.58$ and $f_{R_{1}}=f_{R_{2}}=0.10$.}
	\label{figapp:Hex}
\end{figure}

\newpage

\bibliography{ref.bib}